\documentclass[final, 3p,times]{elsarticle}

\usepackage{graphics}
\usepackage{graphicx}
\usepackage{epsfig}
 
\usepackage{amssymb}
\usepackage{amsthm}

\usepackage{lineno}

\usepackage{color}

\begin{document}

\begin{frontmatter}




\title{Finite scale local Lyapunov exponents distribution in fully developed homogeneous isotropic turbulence}


\author{Nicola de Divitiis}

\address{"La Sapienza" University, Dipartimento di Ingegneria Meccanica e 
Aerospaziale, Via Eudossiana, 18, 00184 Rome, Italy, \\
phone: +39--0644585268, \ \ fax: +39--0644585750, \\
e-mail: n.dedivitiis@gmail.com, \ \  nicola.dedivitiis@uniroma1.it}


\begin{abstract}
The present work analyzes the distribution function of the finite scale local Lyapunov exponent of a pair fluid particles trajectories in fully developed incompressible homogeneous isotropic turbulence.
According to the hypothesis of fully developed chaos, this PDF is reasonably estimated by maximizing the entropy associated to such distribution, resulting to be an uniform distribution function in a proper interval of variation of the local Lyapunov exponents. 
From this PDF, we determine the relationship between the average and maximum Lyapunov exponents 
and the longitudinal velocity correlation function. This link, which leads to the closure of von
K\'arm\'an-–Howarth and Corrsin equations, agrees with the relation obtained in the previous work \cite{deDivitiis_1}, supporting the proposed PDF calculation, at least for the purposes of the energy cascade effect estimation.
Furthermore, through the property that the Lyapunov vectors tend to align to the direction of the maximum growth rate of trajectories distance, we obtain the link between maximum and average Lyapunov exponents in line with the previous result.
\end{abstract}


\end{frontmatter}
 
\newcommand{\no}{\noindent}
\newcommand{\be}{\begin{equation}}
\newcommand{\ee}{\end{equation}}
\newcommand{\bea}{\begin{eqnarray}}
\newcommand{\eea}{\end{eqnarray}}
\newcommand{\bc}{\begin{center}}
\newcommand{\ec}{\end{center}}

\newcommand{\calr}{{\cal R}}
\newcommand{\calv}{{\cal V}}

\newcommand{\bff}{\mbox{\boldmath $f$}}
\newcommand{\bfg}{\mbox{\boldmath $g$}}
\newcommand{\bfh}{\mbox{\boldmath $h$}}
\newcommand{\bfi}{\mbox{\boldmath $i$}}
\newcommand{\bfm}{\mbox{\boldmath $m$}}
\newcommand{\bfp}{\mbox{\boldmath $p$}}
\newcommand{\bfr}{\mbox{\boldmath $r$}}
\newcommand{\bfu}{\mbox{\boldmath $u$}}
\newcommand{\bfv}{\mbox{\boldmath $v$}}
\newcommand{\bfx}{\mbox{\boldmath $x$}}
\newcommand{\bfy}{\mbox{\boldmath $y$}}
\newcommand{\bfw}{\mbox{\boldmath $w$}}
\newcommand{\bfk}{\mbox{\boldmath $\kappa$}}

\newcommand{\bfA}{\mbox{\boldmath $A$}}
\newcommand{\bfD}{\mbox{\boldmath $D$}}
\newcommand{\bfI}{\mbox{\boldmath $I$}}
\newcommand{\bfL}{\mbox{\boldmath $L$}}
\newcommand{\bfM}{\mbox{\boldmath $M$}}
\newcommand{\bfS}{\mbox{\boldmath $S$}}
\newcommand{\bfT}{\mbox{\boldmath $T$}}
\newcommand{\bfU}{\mbox{\boldmath $U$}}
\newcommand{\bfX}{\mbox{\boldmath $X$}}
\newcommand{\bfY}{\mbox{\boldmath $Y$}}
\newcommand{\bfK}{\mbox{\boldmath $K$}}
\newcommand{\bfR}{\mbox{\boldmath $R$}}

\newcommand{\bfrho}{\mbox{\boldmath $\varrho$}}
\newcommand{\bfchi}{\mbox{\boldmath $\chi$}}
\newcommand{\bfphi}{\mbox{\boldmath $\phi$}}
\newcommand{\bfPhi}{\mbox{\boldmath $\Phi$}}
\newcommand{\bflambda}{\mbox{\boldmath $\lambda$}}
\newcommand{\bfell}{\mbox{\boldmath $\ell$}}
\newcommand{\bfxi}{\mbox{\boldmath $\xi$}}
\newcommand{\bfeta}{\mbox{\boldmath $\eta$}}
\newcommand{\bfLambda}{\mbox{\boldmath $\Lambda$}}
\newcommand{\bfPsi}{\mbox{\boldmath $\Psi$}}
\newcommand{\bfXi}{\mbox{\boldmath $\Xi$}}
\newcommand{\bfomega}{\mbox{\boldmath $\omega$}}
\newcommand{\bfOmega}{\mbox{\boldmath $\Omega$}}
\newcommand{\bfeps}{\mbox{\boldmath $\varepsilon$}}
\newcommand{\bfepsn}{\mbox{\boldmath $\epsilon$}}
\newcommand{\bftau}{\mbox{\boldmath $\tau$}}
\newcommand{\bfzeta}{\mbox{\boldmath $\zeta$}}
\newcommand{\bfkappa}{\mbox{\boldmath $\kappa$}}
\newcommand{\bfsigma}{\mbox{\boldmath $\sigma$}}
\newcommand{\bftheta}{\mbox{\boldmath  $\vartheta$}}
\newcommand{\bfmu}{\mbox{\boldmath  $\mu$}}
\newcommand{\itPsi}{\mbox{\it $\Psi$}}
\newcommand{\itPhi}{\mbox{\it $\Phi$}}

\newcommand{\bint}{\mbox{ \int{a}{b}} }
\newcommand{\ds}{\displaystyle}
\newcommand{\Sum}{\Large \sum}



\bigskip

\section{Introduction \label{sect1}}

We analyze the distribution function of the finite scale local Lyapunov exponent of the fluid kinematic field in fully developed homogeneous isotropic turbulence for incompressible fluids. This exponent, defined as
\bea
\ds \tilde{\lambda} \equiv \frac{d \ln \rho}{dt} = \frac{{\bf \dot{{\bfxi}}} \cdot {\bfxi}}{{\bfxi} \cdot {\bfxi}} 
\label{2}
\eea
is a quantity providing the instantaneous growth rate of the distance $\rho = \sqrt{{\bfxi}\cdot {\bfxi}}$ between the two fluid particles trajectories ${\bfx}(t)$ and ${\bfy}(t)={\bfx}(t)+{\bfxi}(t)$, where $\bfxi$ is the separation vector (finite scale Lyapunov vector).
Because of non--smooth spatial variations of the velocity field, $\tilde{\lambda}$ can exhibit fluctuations of sizable amplitude with respect to its average value, thus $\tilde{\lambda}$ plays the role of a stochastic variable and will be distributed according to a certain PDF.

The present analysis is based on the property that the kinematics of a pair of fluid particles, characterized by $\bfxi$, is much faster and statistically independent with respect to the velocity field. 
This property, just discussed in \cite{deDivitiis_1, deDivitiis_2} for what concerns the closure of the von K\'arm\'an-–Howarth and Corrsin equations \cite{Karman38} \cite{Corrsin_1, Corrsin_2}, was previously supported by the arguments presented in Ref. \cite{Ottino90} (and references therein), where the author observes that:
i) the velocity fields ${\bf u} (t, {\bf x})$
produce chaotic trajectories also for relatively simple mathematical structure of the right--hand sides ${\bf u} (t, {\bf x})$. 
ii) the flows given by ${\bf u} (t, {\bf x})$ 
stretch and fold continuously and rapidly causing an effective mixing of the particles trajectories.

Through the hypotheses of fully developed chaos and fluid incompressibility, we first estimate the interval of variation of $\tilde{\lambda}$. Thereafter, due to the fully developed chaos, we
calculate the distribution function of $\tilde{\lambda}$ by maximizing the entropy associated
to such PDF. As the consequence, $\tilde{\lambda}$ results to be uniformely distributed in such interval of variations.
In particular, we determine the link between average and maximum finite scale Lyapunov exponents, $\bar{\lambda}$ and $\lambda$ respectively, resulting $\bar{\lambda}=\lambda/2$.
 These exponents are so defined (see also the Appendix)
\bea
\begin{array}{l@{\hspace{-0.cm}}l}
\ds \bar{\lambda} \equiv \left\langle \tilde{\lambda} \right\rangle \\\\
\label{l ave}
\ds {\lambda} \equiv \left\langle \tilde{\lambda} \right\rangle_{\dot{\xi} \cdot \xi \ge 0} 
\label{L0}
\end{array}
\eea
where  $\ds \left\langle \circ \right\rangle$ and
$\ds \left\langle \circ \right\rangle_{\dot{\xi} \cdot \xi \ge 0}$
denote the average over the entire ensemble of $\bfxi$, and 
the average calculated on the ensemble of Lyapunov vectors where $\ds {\bf \dot{\bfxi}} \cdot \bfxi \ge 0$.

Through such PDF, $\bar{\lambda}$ and $\lambda$ are expressed in terms of the longitudinal velocity correlation function $f = {\left\langle u_r u'_r \right\rangle }/{u^2}$
where
$u^2 = {1}/{3} \left\langle {\bf u} \cdot {\bf u}   \right\rangle$,
$u_r = {\bf u}(t, {\bf x}) \cdot {{\bf r}}/{r}$ and 
$u_r' = {\bf u}(t, {\bf x}+{\bf r}) \cdot {{\bf r}}/{r}$.
This relationship, which leads to the closure formulas of von K\'arm\'an--Howarth and Corrsin equations, coincides with that just presented in Ref. \cite{deDivitiis_1} where the author 
adopts only the average and maximum Lyapunov exponents, properly defined, without considering the distribution of the local exponents. 
Therefore, such PDF should adequately describe the statistics of $\tilde{\lambda}$, 
at least for what concerns the estimation of the energy cascade effects.

Finally, to obtain further confirmation of the previous result, we show that $\lambda = 2 \bar{\lambda}$ for incompressible fluid using the property that the Lyapunov vectors tend to align to the direction of the maximum growth rate of $\ln \rho$ \cite{Ott2002}.

\bigskip

\section{Set of variations of finite scale local Lyapunov exponents in incompressible turbulence \label{sect2}}

This section proposes an analysis for estimating the range of variations of 
$\tilde{\lambda}$ based on the hypotheses of fluid incompressibility and fully developed chaos.

The present analysis starts from the consideration that the turbulent energy cascade is related to the 
fluid particles trajectories divergence, therefore the relative fluid kinematics plays an important role in the estimation of the properties of such energy cascade \cite{deDivitiis_1, deDivitiis_2}.
The relative kinematics is expressed by finite scale Lyapunov vector $\bfxi$ which satisfies 
\bea
\begin{array}{l@{\hspace{-0.cm}}l}
\ds \dot{{\bfx} } = {\bf u} (t, {\bfx}),   \\\\
\ds \dot{{\bfxi} } = {\bf u} (t, {\bfx}+{\bfxi}) - {\bf u} (t, {\bfx})
\end{array}
\label{1}
\eea
where ${\bf u}={\bf u} (t, {\bfx})$ varies according to the Navier--Stokes equations, and ${\bfx}(t)$ and ${\bfy}(t)={\bfx}(t)+{\bfxi}(t)$ are two fluid particles trajectories. The local divergence between ${\bfx}(t)$ and ${\bfy}(t)$ is quantified by $\tilde{\lambda}$ which, due to the bifurcations of the kinematics field (kinematic bifurcations), exhibits oscillations 
of sizable amplitude with respect to its average value. These bifurcations continuously happen in those points of the physical space where
\bea
\ds \det \left( \nabla  {\bf u} (t, {\bf x}) \right) = 0
\label{bifurcations}
\eea 
Observe that the kinematic bifurcations defined by Eq. (\ref{bifurcations}) are not the Navier--Stokes equations bifurcations (dynamic bifurcation), but arise from these latter \cite{deDivitiis_2}. In fact, the Navier--Stokes bifurcations frequentely occur determining continuously non--smooth spatial variations of ${\bf u} (t, {\bf x})$ which in turn lead to the condition (\ref{bifurcations}) in the several points of the space.

The definition of $\tilde{\lambda}$ given by Eq. (\ref{2})
implies that ${\bfxi}$ can be locally expressed as 
\bea
\begin{array}{l@{\hspace{-0.cm}}l}
\ds {\bfxi} = {\bf Q} (t) {\bfxi}(0) \exp(\tilde{\lambda} t), 
\end{array}
\label{3}
\eea
where $\tilde{\lambda}$ plays the role of the stochastic variable and
${\bf Q} (t)$ is a proper orthogonal matrix providing the orientation of ${\bfxi}$ in the
inertial frame $\cal R$. Accordingly, ${\bf \dot{\bfxi}}$ is
\bea
\begin{array}{l@{\hspace{-0.cm}}l}
\ds {\bf \dot{\bfxi}} = \tilde{\lambda} \ {\bfxi} + \bfomega \times {\bfxi}
\end{array}
\label{3'}
\eea
in which $\bfomega$ defines the angular velocity of ${\bfxi}$ with respect to $\cal R$. 
Therefore, the finite scale Lyapunov vectors ${\bfxi}_k$, $k=1, 2, 3$ corresponding to three mutual orthogonal directions are locally represented by  
\bea
\begin{array}{l@{\hspace{-0.cm}}l}
\ds {\bfxi}_k = {\bf Q} (t) {\bfxi}_k(0) \exp(\tilde{\lambda}_k t), \\\\ 
\ds {\bf \dot{\bfxi}}_k = \tilde{\lambda_k}{\bfxi}_k + \bfomega \times {\bfxi}_k, 
\end{array}
 \ \ \ \ \ \ \ \ \ k = 1, 2, 3.
\label{4}
\eea
where $\tilde{\lambda}_k$, $k=1, 2, 3$ are the finite scale local Lyapunov exponents associated to the three directions.
The classical local Lyapunov exponents $\tilde{\Lambda}_k$, $k=1, 2, 3$ are defined for ${\bfxi}_k \rightarrow 0$, $k=1, 2, 3$. 

Now, in order to estimate the set of variations of $\tilde{\lambda}$, observe that,
due to fluid incompressibility, $\nabla \cdot {\bf u} \equiv 0$, and the classical local exponents obey to the following condition
\bea
\begin{array}{l@{\hspace{-0.cm}}l}
\ds \tilde{\Lambda}_1+ \tilde{\Lambda}_2 + \tilde{\Lambda}_3=0, \ \ \tilde{\Lambda}_1 \ge \tilde{\Lambda}_2 \ge \tilde{\Lambda}_3.
\end{array}
\label{5}
\eea
In general, Eq.(\ref{5}) does not hold for finite scale Lyapunov vectors.
Nevertheless, Eq. (\ref{5}) is valid for those  finite scale local exponents 
for which ${\bfxi}_i \times {\bfxi}_j \cdot {\bfxi}_k$ 
is locally preserved for $i \ne j, \ j \ne k, \ k \ne i$, therefore, without lack of generality, such these exponents can be written in the form
\bea
\begin{array}{l@{\hspace{-0.cm}}l}
\ds \tilde{\lambda}_1 = \lambda_m \cos\left( \varepsilon \right), \\\\
\ds \tilde{\lambda}_2 = \lambda_m \cos\left( \varepsilon + \frac{2}{3} \pi \right), \\\\
\ds \tilde{\lambda}_3 = \lambda_m \cos\left( \varepsilon + \frac{4}{3} \pi\right)
\end{array}
\label{iso lambda}
\eea
where $\varepsilon$ and $\lambda_m$ are proper fluctuating quantities depending upon
the velocity field, and
\bea
\begin{array}{l@{\hspace{-0.cm}}l}
\ds \tilde{\lambda}_1+ \tilde{\lambda}_2 +\tilde{\lambda}_3=0, \ \ \tilde{\lambda}_1 \ge \tilde{\lambda}_2 \ge \tilde{\lambda}_3.
\end{array}
\label{6}
\eea
Now, following the hypothesis of fully developed chaos, it is reasonable that $\tilde{\lambda}$ ranges in the set $(\lambda_0, \lambda_S)$, where $\lambda_S>0$ assumes its maximum value
compatible with Eq. (\ref{iso lambda}), and $\lambda_0$ is consequentely calculated.
This implies that 
$\ds \lambda_0 = - \lambda_S/2$, that is
\bea
\begin{array}{l@{\hspace{-0.cm}}l}
\ds \tilde{\lambda} \in \left( -\frac{{\lambda_S}}{2}, {\lambda_S}\right), \\\\
\ds \lambda_S = \sup\left\lbrace \lambda_m \right\rbrace  
\end{array}
\eea

\bigskip

\section{Incompressible fully developed turbulence \label{sect3}}

This section studies the distribution function of $\tilde{\lambda}$,  $P_\lambda=P_\lambda(\tilde{\lambda})$, in fully developed turbulence, where $\tilde{\lambda} \in (-{\lambda}_S/2, {\lambda}_S)$. 
$P_\lambda$ is formally given in terms of the PDF of the Lyapunov vectors, $P$, through the Frobenius Perron equation (see Appendix).

We will show that the proposed estimation of $P_\lambda$ leads to the following relations
\bea
\begin{array}{l@{\hspace{-0.cm}}l}
\ds \langle \tilde{\lambda} \rangle =\frac{\lambda}{2},
\end{array}
\label{eq0}
\eea
\bea
\begin{array}{l@{\hspace{-0.cm}}l}
\ds \left\langle  \tilde{\lambda}^2 \right\rangle = \lambda^2.
\end{array}
\eea
and to the link between $\lambda$ and the longitudinal velocity correlation function.

In order to determine $P_\lambda$, observe that the regime of fully developed turbulence corresponds to a situation of maximum chaos where the kinematic bifurcations cause a total loss of the initial condition data of $\bfxi$. 
Accordingly, it is reasonable that the entropy $\sigma$ associated to $P_\lambda$ assumes its   maximum value compatible with the condition that the integral of $P_\lambda$ over $(-{\lambda}_S/2, {\lambda}_S)$ is equal to the unity, i.e.
\bea
\begin{array}{l@{\hspace{-0.cm}}l}
\ds \sigma = -\int_{-{\lambda}_S/2}^{{\lambda}_S} P_\lambda \ln  P_\lambda  \ d \tilde{\lambda} = \max
\end{array}
\label{smax b}
\eea
\bea
\begin{array}{l@{\hspace{-0.cm}}l}
\ds \int_{-{\lambda}_S/2}^{{\lambda}_S} P_\lambda \ d \tilde{\lambda}  = 1, 
\end{array}
\label{intP}
\eea
Equations (\ref{smax b}) and (\ref{intP}) correspond to the following variational problem
with fixed boundaries
\bea
\begin{array}{l@{\hspace{-0.cm}}l}
\ds J =\int_{-{\lambda}_S/2}^{{\lambda}_S}  \mathfrak{L} \ d \tilde{\lambda} = \max, \\\\
\mathfrak{L} = - P_\lambda \ln  P_\lambda 
+  \eta P_\lambda 
\end{array}
\eea
where $\mathfrak{L}$ and $\eta$ are, respectively, the lagrangian of the problem
and the Lagrange multiplier associated to the condition (\ref{intP}).
The maximum of $\sigma$ is then calculated through steady condition for $J$ ($\delta J =0$),
\bea
\begin{array}{l@{\hspace{-0.cm}}l}
\ds \delta J = \int_{-{\lambda}_S/2}^{{\lambda}_S} 
\frac{\partial \mathfrak{L} }{\partial {P_\lambda}} \ \delta P_\lambda 
  \ d \tilde{\lambda} =0, \ \ \
\forall \ \delta P_\lambda, \  \mbox{compatible with Eq. (\ref{intP})}
\end{array}
\eea 
Because of the arbitrary nature of $\delta P_\lambda$, the maximum of $J$ is achieved for 
\bea
\begin{array}{l@{\hspace{-0.cm}}l}
\ds \frac{\partial \mathfrak{L}}{\partial {P_\lambda}} =0, 
\end{array}
\label{maxc}
\eea
that is  $\tilde{\lambda}$ is uniformely distributed in 
$(-{\lambda}_S/2, {\lambda}_S)$, being
\bea
\ds P_\lambda = 
\left\lbrace 
\begin{array}{l@{\hspace{-0.cm}}l}
\ds \frac{2}{3}\frac{1}{{\lambda}_S}, \ \ \mbox{if} \  \tilde{\lambda} \in \left( -\frac{{\lambda}_S}{2}, {\lambda}_S\right)  \\\\
\ds 0 \ \ \mbox{elsewhere} 
\end{array}\right. 
\label{Pl}
\eea
Hence, $\bar{\lambda}$ and $\lambda$ are calculated in terms of $P_\lambda$ with
Eqs. (\ref{l ave}) and (\ref{l +})  (see appendix)
\bea
\begin{array}{l@{\hspace{-0.cm}}l}
\ds \bar{\lambda} \equiv \left\langle \tilde{\lambda} \right\rangle   = \frac{{\lambda}_S}{4} >0, \\\\
\ds \lambda \equiv \left\langle \tilde{\lambda}  \right\rangle_{\dot{\xi} \cdot \xi \ge 0} = \frac{{\lambda}_S}{2} = 2 \left\langle \tilde{\lambda}\right\rangle
\end{array}
\label{back}
\eea
Next, it is usefull to calculate the mean square $\langle \tilde{\lambda}^2 \rangle$  
\bea
\begin{array}{l@{\hspace{-0.cm}}l}
\ds \left\langle  \tilde{\lambda}^2 \right\rangle = \int_{-{\lambda}_S/2}^{{\lambda}_S}
P_\lambda \tilde{\lambda}^2 \ d \tilde{\lambda} = \lambda^2.
\end{array}
\label{lambda^2}
\eea
According to Eq. (\ref{lambda^2}), the mean square of $\tilde{\lambda}$ equals the square of the average of $\tilde{\lambda}$ calculated for $ {\bf \dot{\bfxi}} \cdot {\bfxi} \ge 0$.

\bigskip

\section{Lyapunov exponents in terms of longitudinal velocity correlation \label{sect4}}

The previous analysis allows to achieve the link between $\lambda$ and the longitudinal velocity correlation function.
In fact, the standard deviation of longitudinal velocity difference directly depends on $u^2$ and $f$ according to
\bea
\begin{array}{l@{\hspace{-0.cm}}l}
\ds  \left\langle (u_r'-u_r)^2 \right\rangle = 2 u^2 \left( 1- f(r) \right) 
\end{array} 
\eea 
On the other hand, the Lyapunov theory gives the longitudinal velocity difference in terms of $\tilde{\lambda}$
\bea
\begin{array}{l@{\hspace{-0.cm}}l}
\ds u_r'-u_r = 
\tilde{\lambda} \left( {\bfxi} \cdot \frac{\bfxi}{\vert \bfxi \vert}\right)_{\ds {\bfxi}={\bf r}}
\end{array}
\eea
Taking into account the isotropy and Eq. (\ref{lambda^2}), $\langle (u_r'-u_r)^2 \rangle$ 
reads as 
\bea
\begin{array}{l@{\hspace{-0.cm}}l}
\ds \left\langle (u_r'-u_r)^2 \right\rangle \equiv \left\langle  \dot{r}^2 \right\rangle  =
 \left\langle \tilde{\lambda}^2  \right\rangle r^2 = \lambda^2 r^2,
\end{array} 
\eea
thus,  $\lambda= \lambda(r)$,  $\bar{\lambda}=\bar{\lambda}(r)$ depend upon the finite scale $r$ 
through the longitudinal velocity correlation $f$
\bea
\begin{array}{l@{\hspace{-0.cm}}l}
\ds \bar{\lambda} (r)=\frac{\lambda(r)}{2} = \frac{u}{r}\sqrt{\frac{1-f(r)}{2}} 
\end{array} 
\label{eq main'}
\eea 
Equation (\ref{eq main'}) leads to the closure of the von K\'arm\'an and Corrsin equations, 
and coincides with that given in Ref. \cite{deDivitiis_1}.
Unlike Ref. \cite{deDivitiis_1},  Eq. (\ref{eq main'}) 
is here achieved exploiting the shape of the distribution (\ref{Pl}). As the proposed analysis  provides results which are in agreement with the previous ones, the hypothesis $\sigma$ =max seems to be an adequate assumption at least for the purposes of calculation of the energy cascade intensity.

\bigskip

\section{Analysis through alignment property of $\bfxi$ \label{sect3 a}}

An alternative way to achieve Eq. (\ref{eq0}) consists in to exploit the fluid incompressibility, and the alignment property of $\bfxi$, following which the Lyapunov vector tends to align to the direction of maximum growth rate of $\rho$ \cite{Ott2002}.
In line with such alignment property, we consider a proper PDF $P_+= P_+(t, {\bf x}, {\bfxi})$ arising from $P(t, {\bf x}, {\bfxi})$ and exhibiting the same level of information of $P$ at least of higher order terms, and which corresponds to a distribution of $\bfxi$ such that ${\bf \dot{\bfxi}} \cdot {\bfxi} \ge 0$, i.e.
\bea
\begin{array}{l@{\hspace{-0.cm}}l}
\forall ({\bf x}, {\bfxi}) \in \left\lbrace {\bf x}\right\rbrace \times \left\lbrace {\bfxi} \right\rbrace \ \ \vert \ \ \ \ {\bf \dot{\bfxi}} \cdot {\bfxi} \ge 0 \ \  P_+(t, {\bf x}, {\bfxi}) > 0, \\\\
\forall ({\bf x}, {\bfxi}) \in \left\lbrace {\bf x}\right\rbrace \times \left\lbrace {\bfxi} \right\rbrace \ \ \vert \ \ \ \ {\bf \dot{\bfxi}} \cdot {\bfxi} < 0 \ \  P_+(t, {\bf x}, {\bfxi}) = 0, \\\\
 {\cal H}(P_+) = {\cal H}(P) + \mbox{H.O.T.}
\end{array}
\label{prop1}
\eea
where $\cal H$ is the entropy associated to $P$
\bea
{\cal H}(P) = - \int_{\bf x} \int_{\xi} P \ln P d {\bf x} \ d {\bfxi} 
\eea

Now, to estimate $P_+$, we apply the mentioned alignment property of $\bfxi$ \cite{Ott2002}: 
in the absence of variations of $\bf x$, such alignment produces ${\bf \dot{\bfxi}} \cdot {\bfxi} \ge 0, \ \forall ({\bf x}, {\bfxi}) \in \left\lbrace {\bf x}\right\rbrace \times \left\lbrace {\bfxi} \right\rbrace$ in the minimum time $\tau = O(1/{\lambda}_S)$, where
\bea
\begin{array}{l@{\hspace{-0.cm}}l}
\ds \Delta {\bfxi} =  {\bf \dot{\bfxi}}(t) \tau + O(\tau^2)
\end{array}
\eea 
is the changing of $\bfxi$ calculated at ${\bf x}=$ const.
Hence, according to the Lyapunov theory, $\ln \rho$ will exhibit variations that, in proximity of its average value $\left\langle \ln \rho \right\rangle$, read as
\bea
\begin{array}{l@{\hspace{-0.cm}}l}
\ds \frac{d \ln \rho}{dt} = \frac{\left\langle \ln \rho \right\rangle_{t+\tau} - \ln \rho }{\tau}
\end{array}
\label{lnro_t}
\eea 
Therefore, we expect that the following PDF  
\bea
\begin{array}{l@{\hspace{-0.cm}}l}
P_+ \left( t, {\bf x}, {\bfxi}\right)= P\left( t, {\bf x}, {\bfxi} + {\bf \dot{\bfxi}} \tau 
 + O(\tau^2) \right) =
P + \nabla_\xi P \cdot {\bf \dot{\bfxi}} \tau + O(\tau^2)
\end{array}
\eea
represents a distribution function with ${\bf \dot{\bfxi}} \cdot {\bfxi} \ge 0$, where 
$P=P\left( t, {\bf x}, {\bfxi} \right)$ and 
$\nabla_\xi P=\nabla_\xi P\left( t, {\bf x}, {\bfxi} \right)$.
Neglecting the higher order terms, we assume that
\bea
\begin{array}{l@{\hspace{-0.cm}}l}
P_+ = P + \nabla_\xi P \cdot {\bf \dot{\bfxi}} \tau 
\end{array}
\label{P_+}
\eea
This PDF identically satisfies Eqs. (\ref{3 L}) and (\ref{prop1}). In fact, for what regards Eq. (\ref{3 L}), the integral over  
$\left\lbrace {\bf x} \right\rbrace \times \left\lbrace{\bfxi} \right\rbrace$ of the first
term at the R.H.S. of Eq. (\ref{P_+}) is equal to one, whereas the integral of the second one
can be reduced to a proper surface integral of $P$ over 
$\partial \left\lbrace {\bfxi} \right\rbrace$ where $P \equiv 0$ through the second Green's identity and fluid incompressibility, thus this latter identically vanishes.
As far as Eq. (\ref{prop1}) is concerned, the entropy ${\cal H} (P_+)$ is 
\bea
\begin{array}{l@{\hspace{-0.cm}}l}
\ds {\cal H}(P_+) = - \int_{\bf x} \int_{\xi} P_+ \ln P_+ d {\bf x} \ d {\bfxi}  \\\\
\ds ={\cal H}(P) - 
\tau \int_{\bf x} \int_{\xi} \nabla_\xi P \cdot {\bf \dot{\bfxi}} \left(1 + \ln P \right) d {\bf x} \ d {\bfxi} + O(\tau^2) 
\end{array}
\label{entropy_+}
\eea
where the second addend at the R.H.S. of Eq. (\ref{entropy_+}) identically vanishes, as, thanks to the fluid incompressibility, it can be reduced to be a surface integral of $P$ over $\partial \left\lbrace {\bfxi} \right\rbrace$ where $P \equiv 0$. Therefore, the choice of $P_+$ given by Eq. (\ref{P_+}) provides
\bea
\begin{array}{l@{\hspace{-0.cm}}l}
\ds {\cal H}(P_+) = {\cal H}(P) + O(\tau^2) 
\end{array}
\label{entropy_++}
\eea
thus it is adequate to estimate $\lambda$.
Accordingly, this latter is  calculated as
\bea
\begin{array}{l@{\hspace{-0.cm}}l}
\ds {\lambda} =
\ds \int_{\bf x} \int_{\xi} P_+(t, {\bf x}, {\bfxi}) \
\frac{\bf {\dot{{\bfxi}}} \cdot {\bfxi}}{{\bfxi} \cdot {\bfxi}} \ d {\bf x} \ d {\bfxi}
\end{array}
\label{l max 2} 
\eea
Substituting Eq. (\ref{P_+}) and (\ref{lnro_t}) into Eq. (\ref{l max 2}), 
we have
\bea
\begin{array}{l@{\hspace{-0.cm}}l}
\ds {\lambda} = \left\langle \tilde{\lambda} \right\rangle -
\ds \int_{\bf x} \int_{\xi} 
\nabla_{\xi} P \cdot {\bf \dot{\bfxi}} \
\left( \ln \rho - \left\langle \ln \rho \right\rangle \right) 
\ d {\bf x} \ d {\bfxi}
\end{array}
\label{ap1}
\eea
Integrating by parts the second addend and taking into account
the fluid incompressibility and the boundary conditions (\ref{bc}), we obtain
\bea
\ds \int_{\bf x} \int_{\xi} 
\nabla_{\xi} P \cdot {\bf \dot{\bfxi}}
\left( \ln \rho - \left\langle \ln \rho \right\rangle \right) 
\ d {\bf x} \ d {\bfxi} = -
\ds \int_{\bf x} \int_{\xi} P(t, {\bf x}, {\bfxi}) \ 
\frac{{\bf \dot{{\bfxi}}} \cdot {\bfxi}}{{\bfxi} \cdot {\bfxi}} \ d {\bf x} \ d {\bfxi} \equiv - \left\langle \tilde{\lambda} \right\rangle
\eea
Hence
\bea
\begin{array}{l@{\hspace{-0.cm}}l}
\ds \lambda  = 2 \left\langle \tilde{\lambda} \right\rangle
\end{array} 
\eea
in agreement with previous results.

\bigskip

\section{Appendix: Lyapunov exponents and vectors distributions \label{sect2 a}}

In this appendix, the link between distribution functions of Lyapunov vectors and Lyapunov
exponents is recalled. 

In order to express the distribution of $\tilde{\lambda}$ in terms of the statistical properties of the solutions of Eq. (\ref{1}), the distribution function of $\bf x$ and $\bfxi$ is first considered.
\bea
\ds P = P(t, {\bf x}, {\bfxi}).
\eea
This PDF changes with the time according to the Liouville theorem  \cite{Nicolis95} associated to Eqs. (\ref{2}). This theorem, arising from the following relation 
\bea
\begin{array}{l@{\hspace{-0.cm}}l}
\ds \int_{\bf x} \int_{\bf \xi}   P \ d {\bf x} \ d {\bfxi}  = 1, \ \ \ \forall t>0,
\end{array}
\label{3 L}
\eea
and from Eqs. (\ref{2}), provides the evolution equation of $P$ \cite{Nicolis95}
\bea
\begin{array}{l@{\hspace{-0.cm}}l}
\ds \frac{\partial P}{ \partial t} +
 \nabla_{\bf x} \cdot \left( P {\bf \dot{{\bf x}}}\right)+
 \nabla_{\bf \xi} \cdot \left( P {\bf \dot{{\bfxi}}}\right) = 0
\end{array}
\label{liouville L}
\eea
where $\nabla_{\bf x} \cdot \left( \circ \right)$ and $\nabla_{\bf \xi} \cdot \left( \circ \right)$
denote the divergence of $(\circ)$ defined in the spaces $\left\lbrace \bf x\right\rbrace$ and 
$\left\lbrace \bfxi \right\rbrace$ respectively, and $d {\bf x}$ and $d {\bfxi}$ are the elemental volumes in the corresponding spaces. 
Taking into account Eq. (\ref{3 L}), and that the homogeneous isotropic turbulence is defined for unbounded fluid domains, $P$ will satisfy the following boundary condition 
\bea
\begin{array}{l@{\hspace{-0.cm}}l}
\ds P = 0, \ \forall ({\bf x}, {\bfxi}) \in \partial \left\lbrace \left\lbrace {\bf x} \right\rbrace \times \left\lbrace {\bfxi} \right\rbrace \right\rbrace \equiv \partial \left\lbrace {\bf x} \right\rbrace 
\bigcup \partial \left\lbrace {\bfxi} \right\rbrace 
\end{array}
\label{bc}
\eea
To describe the statistics of finite scale Lyapunov vectors and exponents for 3D  flows, the initial condition of $P$ is supposed to be of the kind 
\bea
\begin{array}{l@{\hspace{-0.cm}}l}
\ds P(0, {\bf x}, {\bfxi}) 
= \frac{1}{\varphi^3}
\delta \left( \frac{{\bfxi} - {\bf r}}{\varphi}   \right) 
\end{array}
\eea
where, due to homogeneous flow, $\varphi$ = const. 
Accordingly, the statistical average of an integrable function of $\bf x$ and $\bfxi$, say $\zeta$, is
calculated in terms of $P$ 
\bea
\begin{array}{l@{\hspace{-0.cm}}l}
\ds \left\langle {\zeta} \right\rangle = 
 \int_{\bf x} \int_{\bf \xi}  P \ \zeta \ d  {\bf x} \ d {\bfxi},
\end{array}
\label{ave L}
\eea 
In particular, average and maximum finite scale Lyapunov exponents are  
\bea
\begin{array}{l@{\hspace{-0.cm}}l}
\ds \left\langle \tilde{\lambda} \right\rangle =
 \int_{\bf x} \int_{\bf \xi} P(t, {\bf x}, {\bfxi}) \
\frac{{\bf \dot{{\bfxi}}} \cdot {\bfxi}}{{\bfxi} \cdot {\bfxi}} \ d {\bf x} \ d {\bfxi} 
\equiv
\int_\lambda P_\lambda(t, \tilde{\lambda}) \tilde{\lambda} \ d \tilde{\lambda},
\end{array}
\label{l ave}
\eea
\bea
\begin{array}{l@{\hspace{-0.cm}}l}
\ds {\lambda} =
\frac{ \ds \int_{\bf x} \int_{\dot{\xi} \cdot \xi \ge 0} P(t, {\bf x}, {\bfxi}) \ 
\frac{{\bf \dot{{\bfxi}}} \cdot {\bfxi}}{{\bfxi} \cdot {\bfxi}} \ d {\bf x} \ d {\bfxi}} 
{ \ds \int_{\bf x} \int_{\dot{\xi} \cdot \xi \ge 0} P(t, {\bf x}, {\bfxi}) 
\  d {\bf x} d {\bfxi} } 
=
\frac{ \ds \int_{\tilde{\lambda} \ge 0}  P_\lambda(t, \tilde{\lambda}) \ 
\tilde{\lambda} \ d \tilde{\lambda} } 
{ \ds \int_{\tilde{\lambda} \ge 0} P_\lambda(t, \tilde{\lambda}) \ d  \tilde{\lambda} } 
\end{array}
\label{l +}
\eea
where $P_\lambda(t, \tilde{\lambda})$ is the distribution function of $\tilde{\lambda}$ calculated through $P$, by means of the Frobenius Perron equation
\bea
\begin{array}{l@{\hspace{-0.cm}}l}
\ds P_\lambda(t, \tilde{\lambda}) = \int_{\bf x} \int_{\bf \xi} P(t, {\bf x}, {\bfxi}) \ 
\delta\left( \tilde{\lambda} - \frac{{\bf \dot{{\bfxi}}} \cdot {\bfxi}}{{\bfxi} \cdot {\bfxi}} \right) 
\ d {\bf x} \ d {\bfxi},
\end{array}
\eea

\bigskip

\section{Conclusions}

The distribution function of the finite scale local Lyapunov exponent of the kinematics field was established in homogeneous isotropic turbulence. Based on reasonable assumptions regarding the fully developed chaos and the fluid incompressibility, the shape of such distribution and the range of variations of $\tilde{\lambda}$ are determined. This distribution results to be an uniform function in a proper interval of variations. The results arising from such PDF, in particular the link between $\lambda$ and $\tilde{\lambda}$ and the longitudinal velocity correlation, agree with those presented in Ref. \cite{deDivitiis_1}, and this should support the proposed hypothesis for calculating this PDF. An alternative way to proof the relation between such Lyapunov exponents is also presented, which is based on the alignment property of the Lyapunov vectors.

\bigskip

\section{Competing Interests}

The author declares that there is no conflict of interests regarding the publication of this article.

\bigskip

\section{Acknowledgments}

This work was partially supported by the Italian Ministry for the Universities 
and Scientific and Technological Research (MIUR), and received no specific grant from any funding agency in the public, commercial or not-for-profit sectors.

\bigskip

\bigskip

\end{document}